\documentclass[iop]{emulateapj}
\usepackage{amsmath,txfonts,graphicx}
\newcommand{\f}[1]{\smash{\hat #1}}                
\newcommand{\F}[1]{\boldsymbol{\smash{\hat #1}}}   
\newcommand{\V}[1]{\boldsymbol{#1}}                
\newcommand{\B}[1]{\boldsymbol{\smash{\check #1}}} 
\newcommand{\wA}{\omega_\mathrm{A}}
\newcommand{\rb}{r_\mathrm{b}}

\begin{document}

\title{On Hydromagnetic Stresses in Accretion Disk Boundary Layers}

\author{Martin~E. Pessah$^{1}$ and Chi-kwan Chan$^{2}$}

\affil{$^{1}$Niels Bohr International Academy, Niels Bohr
  Institute, Blegdamsvej 17, 2100 Copenhagen \O, Denmark;
  \texttt{mpessah@nbi.dk} \\
  $^{2}$NORDITA, Roslagstullsbacken 23, 106 91 Stockholm, Sweden;
  \texttt{ckch@nordita.org}}

\shorttitle{Hydromagnetic Stresses in Accretion Disk Boundary Layers}

\shortauthors{Pessah \& Chan}

\begin{abstract}
Detailed calculations of the physical structure of accretion disk
boundary layers, and thus their inferred observational properties,
rely on the assumption that angular momentum transport is opposite to
the radial angular frequency gradient of the disk. The standard model
for turbulent shear viscosity satisfies this assumption by
construction. However, this behavior is not supported by numerical
simulations of turbulent magnetohydrodynamic (MHD) accretion disks,
which show that angular momentum transport driven by the
magnetorotational instability (MRI) is inefficient in disk regions
where, as expected in boundary layers, the angular frequency increases
with radius. In order to shed light into physically viable mechanisms
for angular momentum transport in this inner disk region, we examine
the generation of hydromagnetic stresses and energy density in
differentially rotating backgrounds with angular frequencies that
increase outward in the shearing-sheet framework. We isolate the
modes that are unrelated to the standard MRI and provide analytic
solutions for the long-term evolution of the resulting shearing MHD
waves. We show that, although the energy density of these waves can be
amplified significantly, their associated stresses oscillate around
zero, rendering them an inefficient mechanism to transport significant
angular momentum (inward). These findings are consistent with the
results obtained in numerical simulations of MHD accretion disk
boundary layers and challenge the standard assumption of efficient
angular momentum transport in the inner disk regions. This suggests
that the detailed structure of turbulent MHD accretion disk boundary
layers could differ appreciably from those derived within the standard
framework of turbulent shear viscosity.
\end{abstract}


\section{Introduction}

Basic arguments suggest that the angular frequency $\Omega(r)$ of an
accretion disk surrounding a weakly magnetized star must attain a
maximum value, $\Omega_{\rm max} \equiv \Omega(\rb)$, and decrease
inward \citep[or at least remain constant;
  see][]{2004ApJ...613..506M} to match the angular frequency of the
star at the stellar radius $\Omega_\star(r_\star)$; see, e.g.,
\cite{2002apa..book.....F, 2009apsf.book.....H, 2010apf..book.....A}.
The inner disk region, where $r < \rb$ and $d\Omega/dr \ge 0$, is
referred to as the accretion disk \emph{boundary layer}.
Standard accretion disk theory \citep{1973A&A....24..337S} predicts
that half of the energy released in the accretion process takes place
in this region, estimated to be a fraction of the stellar radius.
The spectrum of the radiated energy depends on the detailed properties
of this layer \citep{1993Natur.362..820N, 1995ApJ...442..337P,
  1996ApJ...473..422P, 2001ApJ...547..355P}; thus understanding the
various processes that determine its properties \citep[see,
  e.g.,][]{2004ApJ...610..977P, 2009ApJ...702.1536B,
  2010AstL...36..848I} is of fundamental importance.

Most detailed calculations for determining the structure of the
boundary layer rely on effective models for turbulent angular momentum
transport. These models are usually built as a turbulent version of the Newtonian
viscous stress between fluid layers in a differentially rotating
laminar flow \citep{1959flme.book.....L}, and thus assume a linear
relationship between the stress and the angular frequency gradient
\citep{1974MNRAS.168..603L}.
This assumption, however, seems at odds with the properties of
magnetohydrodynamic (MHD) turbulence revealed by numerical simulations
of accretion disks \citep{1996MNRAS.281L..21A, 2002MNRAS.330..895A,
  2002ApJ...571..413S, 2008MNRAS.383..683P} which show that angular
momentum transport is inefficient in regions of the disk where
$d\Omega/dr > 0$, which are stable to the standard magnetorotational
instability \citep[MRI; see][]{1991ApJ...376..214B,
  1998RvMP...70....1B}.

Motivated by the need of a deeper understanding of the behavior of an
MHD fluid in a differentially rotating background that deviates from a
Keplerian profile, we study the dynamics of MHD waves in
configurations that are stable to the standard MRI.
Employing the shearing-sheet framework, we show that transient
amplification of shearing MHD waves can generate magnetic energy
without leading to a substantial generation of hydromagnetic stresses.
We discuss the implications of these findings.

\section{Assumptions and Local Model for MHD Disk}

We focus our attention on a subsonic, weakly magnetized fluid for
which the ram pressure and magnetic pressure remain small compared to
their thermal counterpart.
As a first approximation, we thus consider a differentially rotating
fluid with angular frequency $\V{\Omega} = \Omega(r) \B{z}$ and
constant background density $\rho_0$.
This is a reasonable assumption in light of the results presented by
\cite{2002MNRAS.330..895A} and \cite{2002ApJ...571..413S}, who carried
out numerical simulations of boundary layers of unstratified accretion
disks and found that the density fluctuations throughout the simulations
are in general quite small.

We work in the framework of the shearing-sheet approximation
\citep{Hill1878, 1978ApJ...222..850G, 1987MNRAS.228....1N,
  1995ApJ...440..742H}, where the equations describing an
incompressible MHD fluid in a corotating frame are given by
\begin{align}
  \partial_t \V{v} + (\V{v}\cdot\nabla)\V{v}
  &= - 2\Omega_0\B{z}\times\V{v} + 2 q\Omega_0^2 x\B{x} 
     - \smash{\frac{\nabla P}{\rho_0}} \nonumber\\
  &+ \frac{(\V{B}\cdot\nabla)\V{B}}{4\pi\rho_0} + \nu\nabla^2\V{v}
     \,, \label{eq:v}\\
  \partial_t \V{B} + (\V{v}\cdot\nabla)\V{B}
  &= (\V{B}\cdot\nabla)\V{v} + \eta\nabla^2\V{B} \,. \label{eq:B}
\end{align}
Here, $\V{v}(\V{x},t)$ and $\V{B}(\V{x},t)$, with $\nabla\cdot\V{v} =
\nabla\cdot\V{B} = 0$, stand for the velocity and magnetic fields;
$\nu$ and $\eta$ denote the kinematic viscosity and resistivity; and
$\Omega_0 \equiv \Omega(r_0)$ is the corotating angular frequency at a
fiducial radius $r_0$.
The first and second terms on the right hand side of
equation~(\ref{eq:v}) correspond to the Coriolis and tidal forces,
respectively. The local pressure $P$ can be found by the
divergence-less condition of the velocity field.
Recalling that the local density $\rho_0$ is assumed to be constant,
and in order to simplify notation, hereafter we redefine the symbols
denoting the pressure and magnetic field in such a way that and
$P/\rho_0 \rightarrow P$ and $\V{B}/(4\pi\rho_0)^{1/2} \rightarrow
\V{B}$.

We decompose the flow into mean and fluctuations as
\begin{align}
  \V{v}(\V{x}, t) &\equiv \V{U}_1(x) + \V{u}(\V{x}, t), \label{eq:v_decomp}\\
  \V{B}(\V{x}, t) &\equiv \V{B}_0(t) + \V{b}(\V{x}, t). \label{eq:B_decomp}
\end{align}
The leading order background velocity is $\V{U}_1(x) \equiv
-q\,\Omega_0 x \B{y}$, where the shear parameter $q$ is given by
\begin{align}
  q \equiv \left.-\frac{d\ln \Omega}{d\ln r}\right|_{r_0} \,.
\end{align}
The homogeneous background magnetic field is in general a function of
time and it evolves according to the induction equation~(\ref{eq:B}),
i.e., $\partial_t\V{B}_0 = - q\,\Omega_0 B_{0x} \B{y}$.

The substitution of equations~(\ref{eq:v_decomp}) and
(\ref{eq:B_decomp}) into (\ref{eq:v}) and (\ref{eq:B}) leads to a
non-linear system for the dynamical evolution of the perturbations
$\V{u}(\V{x}, t)$ and $\V{b}(\V{x}, t)$.
However, as pointed out in \citet{1994ApJ...432..213G}, all the
non-linear terms in the resulting equations vanish identically if we
consider the evolution of a single Fourier mode\footnote{Formally
  speaking we consider two modes, with wavenumbers $\V{k}$ and $-\V{k}$.
  However, because the functions under consideration are real, the
  Fourier coefficients satisfy $\f{f}_{-\V{k}} = \f{f}_{\V{k}}^*$.}.
In this case we obtain
\begin{align}
  \mathcal{D}_t\V{u} - q\,\Omega_0 u_x \B{y}
  &= \V{B}_0\cdot\nabla\V{b} + \nu \nabla^2\V{u}
  - 2 \Omega_0 \B{z}\times\V{u} - \nabla P, \label{eq:u}\\
  \mathcal{D}_t\V{b} + q\,\Omega_0 b_x \B{y}
  &= \V{B}_0\cdot\nabla\V{u} + \eta\nabla^2\V{b}. \label{eq:b}
\end{align}
The ``semi-Lagrangian'' time derivative $\mathcal{D}_t \equiv
\partial_t + \V{U}_1\cdot\nabla$ accounts for advection by the
shearing background.
The shearing component in the Coriolis term cancels out the tidal
force.
We remark that equations~(\ref{eq:u}) and (\ref{eq:b}) are \emph{not}
just linearized equations, they remain valid even if the amplitude of
the perturbations is not small compared to the background values, and
they are exact as long as a single Fourier mode is considered.
Under these conditions, it is sensible to study the evolution of
$\V{u}(\V{x}, t)$ and $\V{b}(\V{x}, t)$ for a long time.

In order to solve equations~(\ref{eq:u}) and (\ref{eq:b}), it is
convenient to work in Fourier space.
The $x$-dependence of the ``semi-Lagrangian'' time derivative can be
removed by employing a shearing coordinate system $(x', y', z', t')
\equiv (x, y + q\,\Omega_0 x t, z, t)$ in which $\mathcal{D}_t =
\partial_{t'}$ \citep{1965MNRAS.130..125G}.
A single mode with a fixed ``shearing'' wavenumber $\V{k}'$ is thus
given by
\begin{align}
  \V{u}(\V{x},t) &= 2\mathrm{Re}
  \left[\F{u}_{\V{k}'}(t)\exp(i\V{k}'\cdot\V{x}')\right],\label{eq:u_fourier}\\
  \V{b}(\V{x},t) &= 2\mathrm{Re}
  \left[\F{b}_{\V{k}'}(t)\exp(i\V{k}'\cdot\V{x}')\right],\label{eq:b_fourier}
\end{align}
where $\V{k}'\cdot\V{x}' = \V{k}(t)\cdot\V{x} = (k_x' + q\,\Omega_0 t
k_y') x + k_y' y + k_z' z$.

Substituting the ansatz~(\ref{eq:u_fourier}) and (\ref{eq:b_fourier}) 
into equations~(\ref{eq:u}) and (\ref{eq:b}) leads a set of equations
for the Fourier amplitudes
\begin{align}
  d_t\F{u} - q\,\Omega_0 \f{u}_x \B{y} &= i\wA\F{b} - \nu k^2\F{u}
  - 2 \Omega_0 \B{z}\times\F{u} - i\V{k} \f{P},
  \label{eq:uk}\\
  d_t\F{b} + q\,\Omega_0 \f{b}_x \B{y} &= i\wA\F{u} - \eta k^2\F{b}.
  \label{eq:bk}
\end{align}
Here, we have replaced $\partial_{t '}$ by $d_t$ and omitted the 
subscripts $\V{k}'$ in the Fourier coefficients in order to 
simplify the notation. We have also introduced the (time-independent) Alfv\'en frequency 
$\wA \equiv \V{B}_0(t)\cdot\V{k}(t)$ (see also \citealt{1992ApJ...400..610B}).

The pressure term can be eliminated from equation~(\ref{eq:uk}) using
the solenoidal character of the velocity field, which implies $d_t
i\V{k} = i\V{k} d_t + iq\,\Omega_0 k_y\B{x}$.
This leads to
\begin{align}
  \f{P} = -\frac{2i\Omega_0}{k^2}\Big[(q-1)k_y\f{u}_x + k_x\f{u}_y\Big],
  \label{eq:pressure_term}
\end{align}
which is independent of $\f{u}_z$.
Because we are interested in the transport of angular momentum along
the radial direction, this decoupling allows us to solve separately
for the $x$- and $y$-components.

\section{Dynamical Evolution of MRI-stable Modes}

\begin{figure*}
   \includegraphics[width=\columnwidth,trim=16 8 8 16]{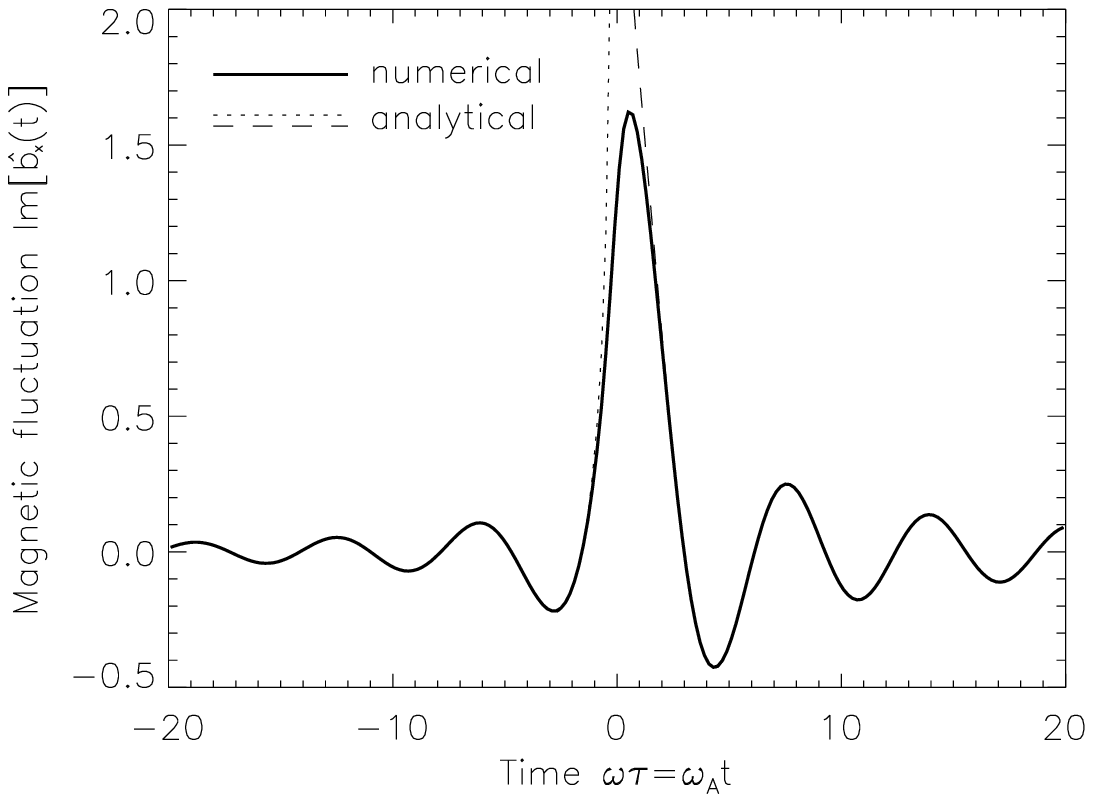}
   \hfill
   \includegraphics[width=\columnwidth,trim=16 8 8 16]{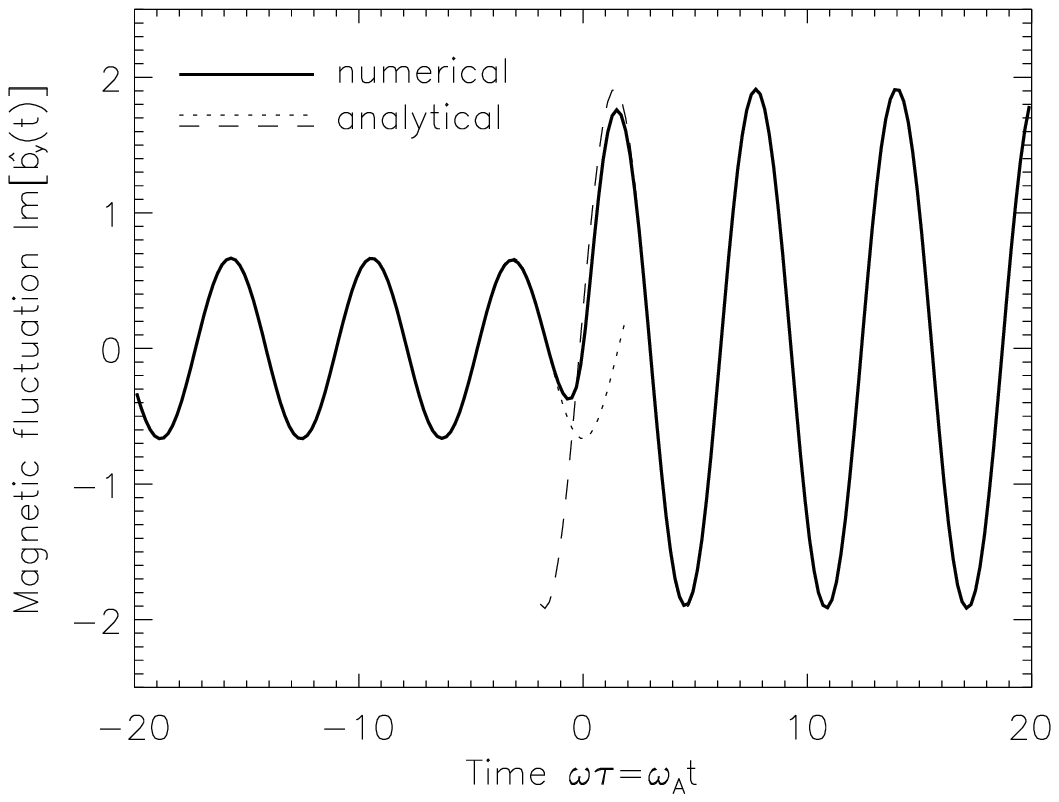}  
   \caption{The Fourier amplitudes for the magnetic field components
     $\hat{b}_x$ (left) and $\hat{b}_y$ (right).
     The thick solid lines are the numerical solutions obtained by
     solving equation~(\ref{eq:2nd}) with $q= -3/2$ and
     $\wA = \Omega_0$ and using the divergence-less conditions~(\ref{eq:divless}).
     The asymptotic solutions for early and late times are both given
     by the analytical approximation~(\ref{app:bx}) and (\ref{app:by})
     but with different integration constants.
     The dotted lines correspond to $C_- = 1$ and $S_- = 0$, which are
     used to specify the initial conditions at $\omega\,\tau = -20$.
     The dashed lines correspond to $C_+ = 0.285$ and $S_+ = -1.896$,
     which are obtained by matching the analytic and numerical results
     at late times, e.g., $\omega\,\tau = 20$.
     Note that, for the sake of clarity, we truncate the analytical
     solutions in the right panel at $\omega\,\tau = 2$ and $-2$.}
  \label{fig:uxbx}
\end{figure*}

The dynamical evolution of the modes with $\V{k} \equiv k_z\B{z}$ is
quite simple; they grow exponentially if $k_z^2 \wA^2 \le 2 q
\Omega_0^2$ \citep{1992ApJ...392..662B, 2006MNRAS.372..183P}.
Thus, a Keplerian disk (with $q=3/2$) can exhibit exponential growth but a shear
profile with $q < 0$ only supports stable oscillations.
In order to isolate the interesting dynamics that could arise from
modes that are not associated with the MRI, we thus focus on modes
with $k_z = 0$.
Taking the curl of the momentum equation, it is easy to verify that
the Coriolis term does not play a role in the equation for the 
vorticity, and hence it has no effect on the dynamics of the system
\citep{2007ApJ...670..789L}.
This shows explicitly that the standard MRI is absent in our analysis.

We choose the origin of time so that $k_x(t)$ is initially zero.
In other words, we use $k_x(t)$ to define our time coordinate,
\begin{align}
  \tau \equiv k_x(t)/k_y \equiv q\,\Omega_0 t \,,
\end{align}
so the divergence-less conditions become
\begin{align}
  \tau\,\f{u}_x + \f{u}_y = \tau\,\f{b}_x + \f{b}_y = 0. \label{eq:divless}
\end{align}
Assuming $\nu, \eta \ll \wA/k^2$, we can work in the ideal limit and
neglect viscosity and resistivity.
The $x$-components of the MHD equations then become
\begin{align}
  d_\tau
  \begin{pmatrix}\f{u}_x \\ \f{b}_x\end{pmatrix} =
  \begin{pmatrix}-\Gamma & i\omega \\ i\omega & 0 \\\end{pmatrix}
  \begin{pmatrix}\f{u}_x \\ \f{b}_x\end{pmatrix}, \label{eq:uxbx}
\end{align}
where all the temporal dependence is contained in the factor
\begin{align}
  \Gamma(\tau) \equiv 2 \tau / (\tau^2 + 1), \label{def:Gamma}
\end{align}
and
\begin{align}
  \omega \equiv \wA / q\,\Omega_0
\end{align}
is the dimensionless Alfv\'en frequency.
The linear system (\ref{eq:uxbx}) can be recast into one equation as
\begin{align}
  d_\tau^2 \f{b}_x + \Gamma(\tau)\,d_\tau \f{b}_x + \omega^2 \f{b}_x = 0,
  \label{eq:2nd}
\end{align}
where the dependence on the parameters $q$, $\Omega_0$, and $\wA$ is
\emph{only} through the combination $(\wA / q\,\Omega_0)^2$.
This second order differential equation for $\f{b}_x$ is identical to
equation (2.20) in \citet{1992ApJ...400..610B} when $k_z=0$.
In this case, the perturbations in the $z$-coordinate decouple from
the perturbations in the perpendicular direction, see also their
equation (2.19).
  
Unfortunately, equation~(\ref{eq:2nd}) does not have an analytical
solution.
However, if we consider the limit $\tau^2 \gg 1$, it reduces to a
spherical Bessel equation, which posseses as solutions
\begin{align}
  \f{u}_x &= S j_1(\omega\,\tau) + C y_1(\omega\,\tau), \nonumber\\
          &= - \frac{C + S \omega\,\tau}{\omega^2 \tau^2}\cos(\omega\,\tau)
             + \frac{S - C \omega\,\tau}{\omega^2 \tau^2}\sin(\omega\,\tau),
  \label{app:ux}\\
  \f{b}_x &= -i S j_0(\omega\,\tau) - i C y_0(\omega\,\tau) \nonumber\\
          &=   \frac{iC}{\omega\,\tau}\cos(\omega\,\tau)
             - \frac{iS}{\omega\,\tau}\sin(\omega\,\tau),
  \label{app:bx}
\end{align}
where $j_n(x)$ and $y_n(x)$, with $n=0,1$, are spherical Bessel
functions of the first and second kind, respectively; and $S$ and $C$
are complex constants determined by the initial conditions.
Using the divergence-less conditions in equation~(\ref{eq:divless}), the
$y$-components are simply
\begin{align}
  \f{u}_y
  &= -   \tau\left[S j_1(\omega\,\tau) + C y_1(\omega\,\tau)\right]\nonumber\\
  &=   \frac{C + S\omega\tau}{\omega^2\tau}\cos(\omega\,\tau)
     - \frac{S - C\omega\tau}{\omega^2\tau}\sin(\omega\,\tau), \label{app:uy}\\
  \f{b}_y
  &=\,\,i\tau\left[S j_0(\omega\,\tau) + C y_0(\omega\,\tau)\right]\nonumber\\
  &= - \frac{iC}{\omega}\cos(\omega\,\tau)
     + \frac{iS}{\omega}\sin(\omega\,\tau). \label{app:by}
\end{align}
Because the pressure in equation~(\ref{eq:pressure_term}) is
independent of $\f{u}_z$, the exact solutions (for all time) are
Alfv\'en waves,
\begin{align}
  \f{u}_z &=\,C'\cos(\omega\,\tau) +\;S'\sin(\omega\,\tau), \label{sol:uz}\\
  \f{b}_z &= iC'\sin(\omega\,\tau) - iS'\cos(\omega\,\tau), \label{sol:bz}
\end{align}
where $C'$ and $S'$ are some other complex constants. We note that,
although the direction of the dimensionless time $\tau$ depends on the
sign of the shear parameter $q$, the combination $\omega\,\tau \equiv
\wA t$ is insensitive to it.
Hence, given the same initial conditions, $\f{u}_x$ and $\f{b}_x$ are
symmetric, while $\f{u}_y$ and $\f{b}_y$ are anti-symmetric, in the
shear parameter $q$.

We demonstrate the accuracy of these analytical approximations in
Figure~\ref{fig:uxbx}, which shows both the numerical and analytical
solutions for $\mathrm{Im}[\f{b}_x]$ and $\mathrm{Im}[\f{b}_y]$ with
$q = -3/2$ and $\wA = \Omega_0$ as an example.
The initial conditions are set at $\omega\,\tau = \wA t = -20$ by
choosing $C_- = 1$ and $S_- = 0$.
The numerical solutions, shown with thick solid lines, result from
integrating equation~(\ref{eq:2nd}) with the
definition~(\ref{def:Gamma}) and using the divergence-less
conditions~(\ref{eq:divless}).
The dotted lines in the two panels are obtained by setting $C = C_-$
and $S = S_-$ in the analytical approximations~(\ref{app:bx}) and
(\ref{app:by}).
These solutions are indistinguishable for $\tau \lesssim -1$.
As expected, the approximations break down for $\tau \simeq 0$.
This is precisely where the numerical solutions change their
amplitudes significantly.
The analytical expressions~(\ref{app:bx}) and (\ref{app:by}) are again
in excellent agreement with the numerical solutions for $\tau \gtrsim
1$, provided that their amplitudes are given by $C = C_+ = 0.285$ and
$S = S_+ = -1.896$.
These constants are found by requiring that both the numerical and
analytical solutions match for $\omega\,\tau = \wA t \gg 1$ (in
practice we set $\omega\,\tau=20$).

Even though our analytical approach cannot predict the change in
amplitude close to $\tau \approx 0$, the solutions that we obtain
are a very good approximation to the numerical results as long as $\tau^2 > 1$.
We could in principle obtain the coefficients $C_+$ and $S_+$ by an
asymptotic matching technique similar to the one employed in
\citet{2009MNRAS.397...52H}.
However, the analytical solution near $\tau \simeq 0$ contains special
functions that are too complicated to be useful.
More importantly, as we show below, the most interesting features of the
solutions are independent of the precise values of these constants.

\begin{figure}
  \includegraphics[width=\columnwidth,trim=16 8 8 16]{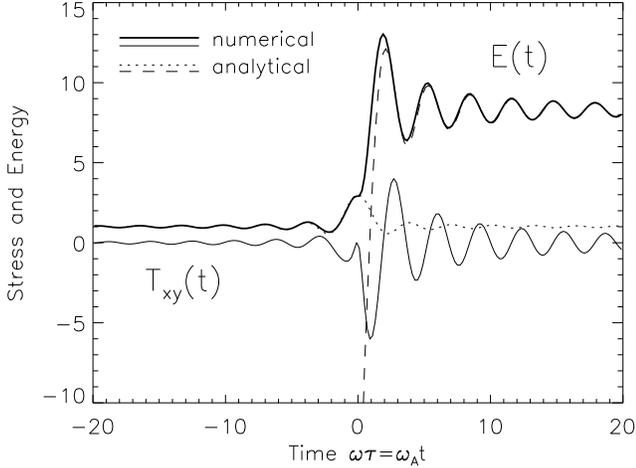}
  \caption{The thick and thin solid lines correspond, respectively, to
    the total energy $E(t)$ and total stress $T_{xy}(t)$, calculated
    using the Fourier amplitudes $\f{u}_x$, $\f{u}_y$, $\f{b}_x$, and
    $\f{b}_y$ obtained numerically.
    The dotted and dashed lines show the analytical approximation for
    the energy, equation~(\ref{eq:E}), using the two sets of constants
    described in the caption of Figure~\ref{fig:uxbx}.
    The $x$-components of the velocity and magnetic fields are
    symmetric in $q$ but the $y$-components are anti-symmetric.
    Therefore, changing the sign of the shear parameter $q$ changes the 
    sign of the stress $T_{xy}$ but not of the energy $E$.}
  \label{fig:se}
\end{figure}

\section{Late Time Stress and Energy}

Given the solutions~(\ref{app:ux}) -- (\ref{sol:bz}) for the Fourier
amplitudes, we obtain the (mean) total stress $\smash{T_{xy}} \equiv
\langle \smash{u_x u_y} - \smash{b_x b_y}\rangle$ and (mean) energy
density $E\equiv \langle \smash{u^2 + b^2} \rangle/2$ of the
fluctuating fields\footnote{We do not include the stress and energy
  generated by the time-dependent mean field $\V{B}_0(t)$ here.
  This contribution depends on the initial magnetic field.}, where the
brackets stand for the spatial average, see, e.g.,
\citet{2006MNRAS.372..183P}.
Because these solutions are only valid for early/late times, we can
approximate the total stress and energy density up to first order in
$1/\omega\,\tau$ as
\begin{align}
  \!T_{xy} \!\approx\!-\frac{2}{\omega^2\tau}
  &\Big[(|S|^2 - |C|^2)\cos(2\omega\,\tau)
      + ( S^*C +  SC^*)\sin(2\omega\,\tau)\Big],
  \label{eq:Txy}\\
  E \!\approx\! -\smash{\frac{1}{\omega^3\tau}}
  &\Big[(|S|^2 - |C|^2)\sin(2\omega\,\tau)
      - ( S^*C +  SC^*)\cos(2\omega\,\tau)\Big]\nonumber\\
  &+ \frac{1}{\omega^2} \left(|S|^2 - |C|^2\right) + |S'|^2 + |C'|^2 \,,
  \label{eq:E}
\end{align}
where the asterisk denotes complex conjugation. 
Using these expressions, it is easy to see that the energy balance 
equation $d_t E = q \Omega_0 T_{xy}$ is also satisfied up to order $1/\omega\,\tau$.

In Figure~\ref{fig:se}, we illustrate the numerical solutions for the stress
$T_{xy}(t)$ and energy $E(t)$, given by the thin and thick solid
lines, together with the analytical approximation for the energy.
The latter has been obtained by substituting the two pairs of
constants, $C_- = 1$ and $S_- = 0$, and $C_+ = 0.285$ and $S_+ =
-1.896$, into equation~(\ref{eq:E}). 
It is thus clear that the late-time stress oscillates around zero with
decreasing amplitude, while the energy density asymptotes
to a non-vanishing, time-independent value.
The expression for the energy density at early/late times in terms of
the constants $S_\pm$ and $C_\pm$ is given by\footnote{Here, we assume
  $S' = C' = 0$, and thus avoid the uninteresting contributions due to
  any stable Alfv\'en wave initially present, see
  equations~(\ref{sol:uz})--(\ref{sol:bz}).}
\begin{align}
  E_\pm \equiv \lim_{t\rightarrow\pm\infty} E(t) =
  \frac{|S_\pm|^2 + |C_\pm|^2}{\omega^2}\,.
  \label{eq:amplification}
\end{align}
Therefore, the energy gain via swing amplification, $E_+/E_-$, is in
general a function of the ratio $\omega = \wA/q\,\Omega_0$ and the
initial conditions.
However, it is possible to obtain conclusions that are independent of
the latter.

\begin{figure}
  \includegraphics[width=\columnwidth,trim=16 8 8 16]{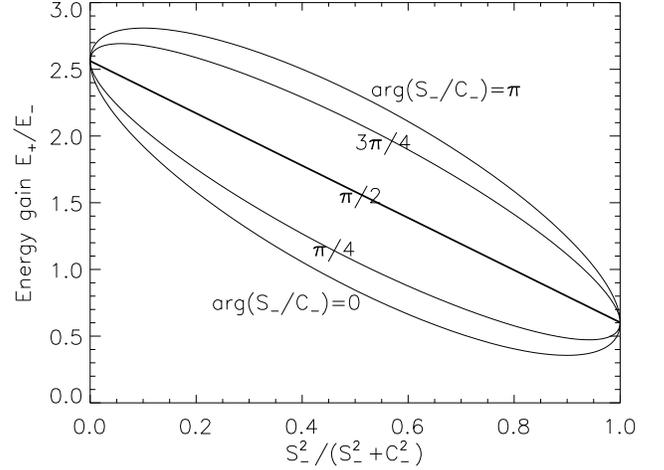}
  \caption{The solid curves show the dependence of the energy gain,
    $E_+/E_-$, on the initial conditions for $\omega = \wA/q\,\Omega_0
    = 1$.
    The horizontal axis provides a measure of the relative amplitude
    of $C_-$ and $S_-$; while the different lines show the results for
    various phase differences.
    For strong shear $1/\omega^2 > 1$, the dependence of the phase is
    weaker and all the different curves collapse onto the straight thick
    line, which corresponds to $\arg(S_-/C_-) = \pi/2$, while this one
    gets steeper, thus providing a larger energy gain, see also
    Figure~\ref{fig:rq}.}
  \label{fig:rs}
\end{figure}

The dependence of the energy gain on the
initial conditions for $\omega^2= 1$ is shown in Figure~\ref{fig:rs}.
The horizontal axis describes how the initial energy is distributed
between the $j_n$ modes and the $y_n$ modes; while the different lines
show the phase difference in the corresponding initial amplitudes.
When $\arg(S_-/C_-) = \pi/2$ or $3\pi/2$, the $j_n$ and $y_n$ modes
are completely out of phase and evolve independently.
This results in an energy gain which is linear in the initial
amplitudes (thick solid line).
We have found that the dependence of the energy gain on the phase
difference between the constants determining the initial conditions is
weaker if $\omega$ decreases below unity.
In this case, all the different curves converge to the thick line
corresponding to $\arg(S_-/C_-) = \pi/2$.
At the same time, as $\omega$ decreases below unity, this line gets
steeper, providing thus a larger energy gain.
This justifies referring to the $y_n$ and $j_n$ as the ``growing'' and
``decaying'' modes, respectively.

We illustrate the dependence of the energy gain on the shear parameter in Figure~\ref{fig:rq}
(because the results depend only on $\omega^2$, we only show the positive domain 
in the horizontal axis). In the limit of weak shear, there are only pure Alfv\'en waves and
there is no net energy gain.
The dashed line shows that the energy gain tends to the value $E_+/E_-
= 10/\omega^2$ as $1/\omega \gg 1$, which provides a good description
of the numerical results for strong shear.
The asymptotic behavior is insensitive to the initial conditions as
long as the growing mode is excited, i.e. $C_-\ne 0$.

\section{Discussion}

\subsection{Summary and Connection to Previous Work}

We have employed the shearing-sheet framework to study the dynamical
evolution of MHD waves in weakly magnetized differentially rotating
backgrounds which are stable to the MRI. While the fact that these 
waves can be transiently amplified is widely appreciated, 
our motivation to study them, as well as the
results that we obtained, concern dynamical aspects that have not
received as much attention.  This is whether these shearing MHD waves 
can play a significant role in the transport of angular momentum in regions of
the disk where the MRI is inefficient, such as the accretion disk boundary layer.

\begin{figure}
  \includegraphics[width=\columnwidth,trim=16 8 8 16]{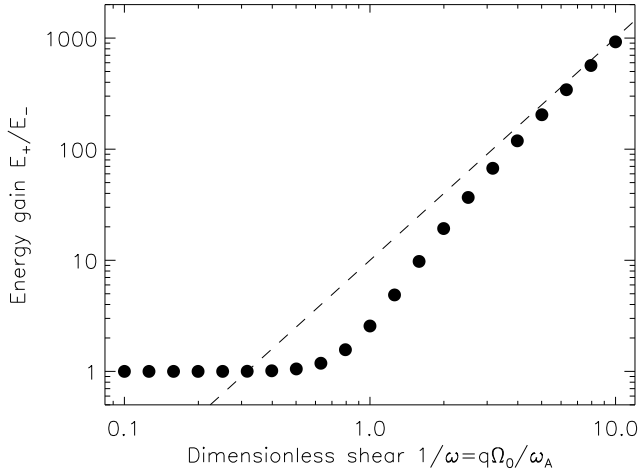}
  \caption{The filled circles represent the value of the energy gain $E_+/E_-$ for
    different values of the shear, parameterized via $q\Omega_0/\wA =
    1/\omega$, obtained via numerical integration using the initial
    conditions $C_- = 1$ and $S_- = 0$, i.e., only the growing mode is
    excited.
    In the limit of weak shear, there are only pure Alfv\'en waves and thus
    there is no net energy gain.
    The dashed line shows the function $10/\omega^2 \equiv
    10(q\Omega_0/\wA)^2$, which is in good agreement with the
    numerical results for strong shear.
    This asymptotic behavior is independent of the initial conditions
    as long as $C_- \ne 0$.}
  \label{fig:rq}
\end{figure}

The equations that we have solved are similar to those presented in
\citet{1992ApJ...400..610B}, who provided numerical solutions showing
that transient amplification of MHD waves is a general outcome for the
modes with wavevectors that are not exactly aligned with the rotation
$z$-axis\footnote{For an analysis of non-axisymmetric spiral waves
  when only a strong, vertical background magnetic field is considered
  see \citet{1992ApJ...393..708T}.}.
Analytical insight on the non-linear dynamics of these waves has been
usually hindered by the fact that the governing equations cannot be
simplified beyond a set of coupled differential equations (see, e.g.,
\citealt{1997MNRAS.291...91F, 2000ApJ...540..372K,
  2006A&A...450..437B}, and also \citealt{2007ApJ...660.1375J}, where
higher order WKB solutions for the linear evolution of the shearing
waves are provided).
In this paper, by isolating the modes that are unrelated to the
standard MRI, i.e., by setting $k_z=0$, we have been able to provide
analytical solutions that are valid for all times, except close to the
instant where the waves evolve from leading to trailing, i.e., when
$k_x(t)=0$.
These solutions are exact when only one Fourier mode is considered
\citep{1994ApJ...432..213G}.

Despite the fact that we do not predict analytically the amplification
factor for these waves, the characteristics of the solutions that we
found allowed us to draw important conclusions.
We showed that the amplification factor is \emph{only} a function of
the (time-independent) dimensionless ratio $(q\Omega_0/\omega_{\rm
  A})^2$, with 
\begin{align}  
  \frac{E_+}{E_-} \approx 10 \left(\frac{q\Omega_0}{\omega_{\rm A}}\right)^2 \quad
  \textrm{for} \quad q\Omega_0 \gg \omega_{\rm A} \,,
\end{align}    
see equation~(\ref{eq:amplification}) and Figure~\ref{fig:rq}, and it is
thus insensitive to the sign of the shear parameter $q$.

An important result of this study is that while the energy of these 
MHD waves can be significantly amplified, their net associated stresses 
oscillate around zero, see equation~(\ref{eq:Txy}) and Figure~\ref{fig:se}.
This suggests that these shearing MHD waves are unlikely to play an important
role in the transport of angular momentum in the accretion disk boundary layer region.
These findings are consistent with the results of global MHD simulations
of accretion disks with a rigid inner boundary carried out in
\citet{2002MNRAS.330..895A} and \citet{2002ApJ...571..413S}. 
These simulations show that the inner disk regions, where $d\Omega/dr \ge 0$, 
can develop strong toroidal magnetic fields, with associated magnetic energies 
that can easily reach a few tenths of the thermal energy, without leading to 
efficient angular momentum transport.

\subsection{Implications}

The importance of understanding the relationship between the stress
and the radial gradient in angular frequency resides in that this
dependence plays a key role when modeling the inner structure of an
accretion disk surrounding a weekly magnetized star (see, e.g.,
\citealt{1995ApJ...442..337P, 1996ApJ...473..422P}).

In the steady state, the constant inward flux of angular momentum at
any given radius $r_0$ is given by $\dot J = \dot M l - 2\pi r_0^2 H
T_{r\phi}$, where $\dot M$ stands for the accretion rate, $l$ is the
specific angular momentum, $H$ is the disk height, and $T_{r\phi}$
(denoted by $T_{xy}$ in our analysis) is the component of the stress responsible for the
flux of azimuthal momentum across the radial direction.
Thus, the angular momentum flux has two contributions: $\dot
J_\mathrm{matter} = \dot M l$, which accounts for the flux of angular
momentum due to mass accretion, and $\dot J_\mathrm{stress} = - 2\pi
r_0^2 H T_{r\phi}$, which accounts for the flux of angular momentum
due to the stress acting on the fluid elements constituting the disk.
Under the reasonable assumption that the disk must be Keplerian 
well beyond the boundary layer, i.e., 
for $r \gg \rb$, and that the stress should vanish in the absence of shear, we
must have $\dot J \equiv \dot M l(\rb)> 0$, i.e., a slowly rotating
star accretes mass {\it and} angular momentum.

In order for this picture to be self-consistent, the stress must
satisfy $T_{r\phi}(r_\star \le r\le \rb) \le 0$.
In the standard accretion disk model this requirement is satisfied by
assuming that the stress is linearly proportional to the local shear
$T_{r\phi} \sim - d\Omega/dr$.
With this model for the stress, and some supplementary assumptions, it
is possible to solve for the radial dependence of $\Omega(r)$ and
determine the structure of the disk (see, e.g.,
\citealt{1991ApJ...370..604P}).
However, this assumption, broadly adopted in the framework of enhanced
turbulent disk viscosity, does not seem to be supported by the modern
paradigm, in which angular momentum transport is due to MHD turbulence
driven by the MRI. Indeed, both local numerical simulations of 
shearing-boxes with non-Keplerian shear profiles 
\citep{2008MNRAS.383..683P, 2009A&A...505..955S} and
global disk simulations with a rigid inner boundary
\citep{2002MNRAS.330..895A, 2002ApJ...571..413S} suggest that angular
momentum transport is inefficient if $d\Omega/dr > 0$.
This suggests that the detailed structure of accretion disk boundary layers 
resulting from the interaction of an MHD disk with a weakly magnetized star 
could differ appreciably from those derived within the standard turbulent shear
viscosity, where the direction of angular momentum transport
is always opposite to the angular frequency gradient.

\subsection{Final Remarks}

It is worth mentioning explicitly that the shearing-sheet framework that
we have employed is inherently limited to address the conditions
expected in the accretion disk boundary layer. For example, the
absence of a hard-inner boundary could prevent Kelvin-Helmholtz
instabilities from operating, see, e.g., \citet{2009ApJ...702.1536B}.
However, these instabilities do not seem to play a predominant role in
the global MHD simulations of \cite{2002MNRAS.330..895A} and
\cite{2002ApJ...571..413S}. Moreover, because we have assumed a constant 
background density, our analysis precludes the possibility 
of buoyant modes or convective instabilities. Whether these instabilities,
and the turbulence they could drive, transport angular
momentum inward or outward in Keplerian disks has been long debated
\citep{1992ApJ...388..438R, 1996ApJ...465..874C, 1996ApJ...464..364S, 2010MNRAS.404L..64L}.
To our knowledge, these convective instabilities have 
not been studied in differentially rotating backgrounds with angular frequencies 
increasing outward; and speculating about their role goes beyond the scope of 
the present work.

In spite of the simplifications of our analytical study, 
the explicit solutions that we have found can provide physical insight and 
help elucidate transport processes in the inner disk regions close to a weakly 
magnetized accreting star. 
The current availability of powerful parallel codes (e.g. \citealt{2008ApJS..178..137S})
which are already being used to study the hydrodynamics of accretion disk boundary 
layers \citep{2011arXiv1112.3102B} holds the promise
that a more detailed understanding of MHD boundary layers will soon 
be possible. 

\acknowledgments

MEP is grateful to the Knud H{\o}jgaard Foundation for its generous support.
CKC is supported by a NORDITA fellowship.
We thank Tobias Heinemann, John Wettlaufer, and Jim Stone for useful discussions.


\end{document}